\documentclass{article}

\usepackage{amsmath,amssymb,bm,color,mathrsfs,graphicx}
\usepackage[normalem]{ulem}
\usepackage{mathtools}
\usepackage{authblk}
\usepackage{xcolor}
\usepackage{graphpap}
\usepackage[UKenglish]{isodate}
\usepackage[left=3cm]{geometry}

\newcommand{\bA}{\mathbf{A}}

\newcommand{\bBo}{\mathbf{\mathring{B}}}
\newcommand{\bBoe}{\bBo_{\texttt{e}}}
\newcommand{\bBoeq}{\bBo_{\!_0}}
\newcommand{\bC}{\mathbf{C}}

\newcommand{\bChooke}{\mathbb{C}}

\newcommand{\bCtr}{\bC_{\texttt{tr}}}
\newcommand{\bD}{\mathbf{D}}
\newcommand{\bE}{\mathbf{E}}

\newcommand{\bF}{\mathbf{F}}
\newcommand{\bFe}{\bF_{\!\texttt{e}}}

\newcommand{\bFo}{\mathbf{\mathring{F}}}
\newcommand{\bFoe}{\bFo_{\!\texttt{e}}}

\newcommand{\bG}{\mathbf{G}}
\newcommand{\bH}{\mathbf{H}}
\newcommand{\bI}{\mathbf{I}}
\newcommand{\bK}{\mathbf{K}}
\newcommand{\bL}{\mathbf{L}}
\newcommand{\bM}{\mathbf{M}}

\newcommand{\bN}{\mathbf{N}}
\newcommand{\bQ}{\mathbf{Q}}

\newcommand{\bS}{\mathbf{S}}
\newcommand{\bT}{\mathbf{T}}

\newcommand{\ba}{\mathbf{a}}
\newcommand{\bb}{\mathbf{b}}

\newcommand{\be}{\mathbf{e}}

\newcommand{\bk}{\mathbf{k}}
\newcommand{\bl}{\,\mathbf{l}}
\newcommand{\bkc}{\mathbf{\hat{k}}}
\newcommand{\bn}{\mathbf{n}}
\newcommand{\br}{\mathbf{r}}
\newcommand{\bu}{\mathbf{u}}
\newcommand{\bv}{\mathbf{v}}
\newcommand{\vs}{v_\textrm{s}}

\newcommand{\bno}{\bn_{_0}}

\newcommand{\bna}{\boldsymbol{\nabla}}
\newcommand{\tsp}{\!\top\!}

\newcommand{\eps}{\epsilon}
\newcommand{\epsO}{\varepsilon_{\bn}}
\renewcommand{\phi}{\varphi}
\newcommand{\tp}{\!\otimes\!}

\newcommand{\divr}{\mathop{\mathrm{div}}}

\newcommand{\skw}{\mathop{\mathrm{skw}}}
\newcommand{\sym}{\mathop{\mathrm{sym}}}
\newcommand{\tr}{\mathop{\mathrm{tr}}}
\newcommand{\trp}{\mathop{\mathrm{tr}_{_{\!\perp}}\!\!}}

\newcommand{\dev}{\mathop{\mathrm{dev}}}
\newcommand{\spn}{\mathop{\mathrm{span}}}
\DeclareMathOperator{\re}{Re}
\DeclareMathOperator{\im}{Im}

\newcommand{\adc}[1]{{\color{black}{#1}}}
\newcommand{\pb}[1]{{\color{black}{#1}}}
\newcommand{\Radc}[1]{{\color{black}#1}}
\newcommand{\RADC}[1]{{\color{black}#1}}
\newcommand  {\aE}  {\alpha}
\newcommand  {\an}  {a}
\newcommand  {\alI} {\alpha_{_1}}
\newcommand  {\alII}{\alpha_{_2}}
\newcommand  {\alO} {\alpha_{_0}}
\newcommand  {\aO}  {\an_{_0}}
\newcommand  {\AO}  {A_{_0}}
\newcommand  {\baI} {\ba_{_1}}
\newcommand  {\baO} {\ba_{_0}}
\newcommand  {\bAO} {\bA_{_0}}
\newcommand  {\bHI} {\bH_{_1}}
\newcommand  {\bHId}{\dot{\bH}_{_1}}
\newcommand  {\bmo} {\mathbf{m}}
\newcommand  {\bMo} {\bM_{_0}}
\newcommand  {\bt}  {\mathbf{t}}
\newcommand  {\bTeq}{\bT_{\!_0}}
\newcommand  {\bTh} {\mathbf{\Theta}}
\newcommand  {\bTI} {\bT_{\!_1}}
\newcommand  {\bvI} {\bv_{\!_1}}
\newcommand  {\bw}  {\mathbf{w}}

\newcommand {\bydef}{\,\raise.07485ex\hbox{:}\kern-.025em\hbox{=}\,}
\newcommand {\defby}{\,=\kern-.345em\hbox{\raise.07485ex\hbox{:}}\,}
\newcommand  {\blI} {\bl_{_1}}
\newcommand  {\blO} {\bl_{_0}}

\newcommand  {\Cels}{^{\circ}\mathrm{C}}

\newcommand  {\DD}  {\mathbb{D}}
\newcommand  {\Dv}  {\mathcal{D}_{\bv}}
\newcommand  {\elI} {\lambda_{_1}}
\newcommand  {\elO} {\lambda_{_0}}

\newcommand  {\kg}  {\mathrm{kg}}

\newcommand  {\m}   {\mathrm{m}}
\newcommand  {\muI} {\mu_{_1}}
\newcommand  {\muO} {\mu_{_0}}

\newcommand  {\pE}  {\pi}
\newcommand  {\ph}  {\hat{p}}
\newcommand  {\piso}{p_{\,\texttt{iso}}}
\newcommand  {\pI}  {p_{_1}}
\newcommand  {\pO}  {p_{_0}}

\newcommand  {\Prt} {\mathscr{P}}
\newcommand  {\PH}  {\mathbb{P}_{_{\!\bH}}}
\newcommand  {\Rlx} {\mathscr{R}}
\newcommand  {\ro}  {\varrho}
\newcommand  {\roI} {\ro_{_1}}
\newcommand  {\roO} {\ro_{_0}}
\newcommand{\sigmap}{\hat{\sigma}_{_{\!+}}\!}
\newcommand {\sigmF}{\sigma_{_{\!\textsf{OF}}}\!}
\newcommand {\sigmh}{\hat{\sigma}}
\newcommand {\sigmp}{\sigma_{_{\!+}}\!}
\newcommand {\sigmr}{\widetilde{\sigma}}
\newcommand  {\sO}  {\sigma_{\texttt{iso}}}
\newcommand  {\SLp} {\textsf{SL}^{\!+}}
\newcommand  {\Sym} {\textsf{Sym}}

\newcommand  {\THR} {\mathrm{T}_{_{\!\bH}}\Rlx}
\newcommand  {\TIR} {\mathrm{T}_{_{\!\bI}}\Rlx}
\newcommand  {\vO}  {v_{_0}}
\newcommand  {\Vo}  {V_{\!_0}}
\title{\textbf{Anisotropic wave propagation \\ in nematic liquid crystals}}

\author[1]{Paolo Biscari%
\thanks{\texttt{paolo.biscari@polimi.it}}}

\author[2]{Antonio DiCarlo}

\author[1]{Stefano S.\,Turzi}

\affil[1]{\small{Dipartimento\,di\,Matematica, Politecnico\,di\,Milano, Piazza\,Leonardo\,da\,Vinci,\,32~~20133\,Milano\,(Italy)}}
\affil[2]{Dipartimento\,di\,Matematica\,\&\,Fisica, Universit\`a Roma\,Tre, Via\,Corrado\,Segre,\,6~~00146\,Roma\,(Italy)}

\cleanlookdateon
\date{\today}
\begin{document}
\maketitle
\begin{abstract}
Despite the fact that quantitative experimental data have been available for more than forty years now, nematoacoustics still poses intriguing theoretical and experimental problems. In this paper, we prove that the main observed features of acoustic wave propagation through a nematic liquid crystal cell -- namely, the \adc{frequency-dependent anisotropy of sound velocity and acoustic attenuation} -- \RADC{may be plausibly explained by a first-gradient continuum theory characterized by a hyperelastic anisotropic response from an evolving relaxed configuration. The latter concept -- new in liquid crystal modeling -- provides the first theoretical explanation of the structural relaxation process hypothesized long ago by Mullen et al.\,\cite{72mull}.} We compare and contrast our proposal with a competing theory where the liquid crystal is modeled as an isotropically compressible, anisotropic second-gradient fluid.
\end{abstract}
%
%
\section{Introduction} \label{sec:intro}
It has long been recognized that sound waves interact with the orientational order of a nematic liquid crystal (NLC), often in a rather subtle way \cite{04kapu}. For instance, strong sound waves impinging on an NLC cell have been observed to induce shear flows which, in turn, perturb the nematic alignment. But also in the linear acoustic regime a variety of interesting effects have been detected. A low intensity ultrasonic wave injected into an NLC cell changes its optical transmission properties, namely, its refractive index. Recently, this acousto-optic effect has attracted renewed attention, due to its potential for application to acoustic imaging \cite{00sand,09sand}. A different manifestation of the coupling between acoustic waves and nematic order is the phenomenon of acoustic generation observed in an NLC cell undergoing  Fr\'eedericksz transitions triggered by an external electric field \cite{99kim}. \par
In this paper we concentrate on the anisotropic propagation of acoustic waves through an NLC cell where the mass density and the nematic field are uniform in the unperturbed state. This phenomenon was studied experimentally more than forty years ago by Mullen, L\"{u}thi and Stephen \cite{72mull}. They found that the speed of sound is maximum when the direction of propagation is along the nematic director, minimum when it is orthogonal to it. The difference is minimal -- a few thousandths of the average sound speed -- and increases with the sound frequency approximately linearly in the range 2--14\,MHz, while being roughly independent of temperature in the range 25--35$\,\Cels$. \par
That most remark\-a\-ble paper, while not attempting to construct a complete theory, did give some valuable hints at a plausible theoretical explanation. To our present purposes, this is the key quote from \cite{72mull}: `The experimental anisotropy in the sound velocity indicates that at finite frequencies a liquid crystal has an anisotropic compressibility. This anisotropy can be explained if at these frequencies a liquid crystal in some respects behaves like a solid [\,\dots]\,\Radc{\cite{MLS}}. However, the elastic constants must have an important frequency dependence [\,\dots]. If this were not the case, it would cost a finite energy to change the shape of a liquid crystal, the volume being kept constant. This is not consistent with our present ideas of the structure of a liquid crystal. The frequency dependence of the elastic constants could arise out of some structural relaxation process in the liquid crystal.' \par
Much more recently, a different line of thought was followed by Selinger and coworkers, who propounded a free-energy density containing a term proportional to $(\nabla\!\ro\cdot\bn)^2$, thus postulating a direct coupling between the spatial gradient of the mass density $\nabla\!\ro$ and the nematic director $\bn$ \cite{02seli,03seli,04seli,05seli}. This idea was subsequently fully developed into a theory of anisotropic Korteweg-like fluids by Virga \cite{09virga}. Because of mass conservation, the mass density is related to the determinant of the deformation gradient. Therefore, the Selinger-Virga hypothesis -- differently from the anisotropic compressibility hypothesis advanced in \cite{72mull} -- establishes a second-gradient theory. Since higher-gradient terms compete with the first-gradient term (the standard isotropic energy density \Radc{function} $\ro\mapsto\!\sO(\ro)$ in the case at hand), they typically represent singular perturbations to the underlying first-gradient theory. Accordingly, their
contribution becomes important only if and where abrupt density changes take place. In fact, the original proposal by Korteweg was meant to model interfacial and capillary forces by resolving density discontinuities into smooth but steep density variations \cite{01kort}. It seems therefore implausible that the mild density undulations occurring in the linear acoustic regime may produce sizeable second-gradient effects. \par
Be that as it may, we here explore an alternative explanation for
the slightly anisotropic propagation of acoustic waves in NLCs, building on the experimental results and the theoretical conjectures of Mullen, L\"{u}thi and Stephen \cite{72mull}. Namely, we construct a first-gradient theory consistent with their experimental observations -- frequency dependence included -- by grafting the concept of evolving relaxed configuration \cite{02dicqui} (or `multiple natural configurations' in Rajagopal and Srinivasa's parlance \cite{04raj}) onto a hyperelastic model accounting for the small anisotropic compressibility of NLCs and the related shear stress. In particular, we identify the dissipative dynamics governing the evolution of the relaxed configuration as the \Radc{`structural relaxation process' explicitly hypothesized, but left altogether undetermined in \cite{72mull}}.\par
Our paper is organized as follows. Sec.\,\ref{sec:elas} is devoted to constructing a hyperelastic first-gradient theory of slightly compressible NLCs. In Sec.\,\ref{sec:elas_non-hyper} we show preliminarily that the elastic response included in the early visco-elastic assumption put forward by Ericksen in his seminal work on anisotropic fluids \cite{60Ericksen} suffices to produce an anisotropic (frequency independent) speed of sound. However, we also prove (Sec.\,\ref{sec:elas_nonhyper}) that such an assumption is incompatible with anisotropic hyperelasticity. We then proceed to construct a stored energy density function capable of representing the solid-like behavior of slightly compressible NLCs at high frequency (Sec.\,\ref{sec:elas_energy} and \ref{sec:elas_special}). On this basis, in Sec.\,\ref{sec:elas_sound} a perturbation analysis is used to obtain a satisfactory dependence of the speed of sound on the angle between the wave vector and the nematic director. In Sec.\,\ref{sec:relax} we introduce and
exploit a concept of nematic relaxation, whereby the isochoric component of the stress-free deformation is quickly ``dragged'' towards the correspondent component of the current configuration, thus allowing the shear stress to relax (with a characteristic time much smaller than the director relaxation time). The equation governing the evolution of the relaxed configuration is given in Sec.\,\ref{sec:relax_eq}. A frequency-dependent speed of sound is finally obtained in Sec.\,\ref{sec:relax_sound}, in good agreement with the experimental data reported in \cite{72mull}. Also the anisotropic attenuation we obtain compares well with the experimental findings of other early authors \cite{70lord}. Sec.\,\ref{sec:discussion} contains a discussion, where our results are put in perspective and further developments are proposed. \Radc{Two appendices complete the paper. In Appendix \ref{app:Ericksen} we prove that the early Ericksen's constitutive assumption for the Cauchy stress in an anisotropic elastic fluid cannot 
be hyperelastic. Appendix \ref{app:hyperel}, besides collecting several computational details of our nonlinear nematic hyperelastic theory, presents \RADC{a thorough comparison of its linearized version and the small-displacement theory hinted at by Mullen et al.\,\cite{72mull}.}}
\section{Nematic elasticity}
\label{sec:elas}
Isothermal conditions are assumed in all what follows. Moreover, we assume that the NLC, uniformly aligned in its unperturbed state, stays so while traversed by the acoustic wave. These assumption reflects the setup of all the cited experimental studies \cite{Note}. To get an idea of~the (small) effects of the removal of the constraint on the nematic texture, \Radc{see \cite{11virga,XXturzi}}. To stress that the director field is uniform and stationary, we shall denote it by $\bno$. The ne\-mat\-ic degrees of freedom being frozen, the governing equations reduce to the mass and force balances:
\begin{equation}
\Dv\ro\, + \ro \divr \bv  = 0\,,
\qquad \divr \bT\,-\ro\,\Dv\bv = 0\,.
\label{eq:balances}
\end{equation}
All fields involved -- mass density $\ro$, translational velocity $\bv$, and Cauchy stress $\bT$ -- are spatial, and $\Dv$ denotes convected time differentiation:
\begin{equation}
\Dv\ro\,\bydef\,\dot{\ro}+(\bna\!\ro)\!\cdot\!\bv\,,
\qquad \Dv\bv\,\bydef\,\dot{\bv}+(\bna\bv)\bv\,,
\label{eq:convected}
\end{equation}
where dotted quantities are partial time derivatives.
\subsection{Ericksen's transversally isotropic fluid}
\label{sec:elas_non-hyper}
The constitutive assumption for the stress put forward by Ericksen \cite{60Ericksen} as early as in 1960, when stripped of the viscous terms and under isothermal conditions, reduces to
\begin{equation}
\bT = -\big(\pE(\ro)\, \bI + \aE(\ro)\,\bno\!\tp \bno\!\big)
\label{eq:T_1}
\end{equation}
(modulo a change of notation; see \cite{TP} for a definition of the tensor product). Both the spherical and the \emph{uniaxial} component are assumed to depend only on the mass density. It is worth rewriting \eqref{eq:T_1} as the sum of a spherical and a \emph{deviatoric} (i.e., traceless) component:
\begin{equation}
\bT = -\left(\pE(\ro)+\tfrac13\,\aE(\ro)\right)\bI -
\aE(\ro)\!\left(\bno\!\tp \bno\!-\tfrac13\,\bI\right),
\label{eq:T_1alt}
\end{equation}
so as to make apparent that $\pE(\ro)$ is not the pressure, which depends also on the anisotropic coefficient $\aE(\ro)$. Adopting \eqref{eq:T_1} transforms the force balance \eqref{eq:balances}$_2$ into
\begin{equation}
\ro\,\Dv\bv = -\bna\pE- (\bno\!\tp \bno)\bna \aE\,.
\label{eq:balance_2}
\end{equation}
Linear acoustic waves are obtained via a regular perturbation expansion of \eqref{eq:balances}$_1$ and \eqref{eq:balance_2} around the unperturbed state:
\begin{equation}
\ro_{\eps} = \roO\!+ \eps\,\roI\!+ o(\eps)\,,\qquad
\bv_{\eps} = \eps\,\bvI\!+ o(\eps)
\label{eq:expan}
\end{equation}
where $\eps$ is a smallness parameter. Assumptions \eqref{eq:expan} entail
\begin{equation}
\begin{aligned}
\pE(\ro_{\eps}) & = \pE(\roO) + \eps\,\pE'(\roO)\roI\!+ o(\eps)\,,
\\[.5ex]
\aE(\ro_{\eps}) & = \aE(\roO) + \eps\,\aE'(\roO)\roI\!+ o(\eps)\,.
\end{aligned}
\label{eq:sviluppo_al_2}
\end{equation}
Since $\bna\!\roO\!=0\,$, the $O(1)$ set of equations is trivially satisfied. The $O(\eps)$ set is comprised of the acoustic equations
\begin{equation}
\dot{\ro}{_{_1}}\!+ \roO \divr \bvI\!= 0\,,\qquad
\roO \dot{\bv}{_{_1}}\!= -\bAO\!\bna\!\roI,
\label{eq:bilancio_massa_4}
\end{equation}
where the acoustic tensor $\bAO\,$is given by
\begin{equation}
\bAO\! = \pE'(\roO) \,\bI +\aE'(\roO) \,\bno\!\tp \bno\,.
\label{eq:tensore_acustico_1}
\end{equation}
Equations \eqref{eq:bilancio_massa_4} entail the anisotropic wave equation
\begin{equation}
\ddot{\ro}{_{_1}}\!- \divr(\bAO\!\bna\!\roI) = 0\,.
\label{eq:onde_2}
\end{equation}
A plane wave
\begin{equation}
\roI\!(\br,t) = A\cos\!\big(\bk\!\cdot\!\br - \omega\,t\big)
\label{eq:plane_wave}
\end{equation}
with\;$A$\;the wave amplitude, $\bk$ the wave vector, its modulus $|\bk|$ the wave number, and $\omega$ the angular frequency, solves \eqref{eq:onde_2} if its phase velocity $\vs\!\bydef\omega/|\bk|$ solves the quadratic equation
\begin{equation}
\vs^2 = \pE'(\roO) + \aE'(\roO) \, (\bk\!\cdot\!\bno/|\bk|)^2.
\label{eq:velfase}
\end{equation}
Common sense and experimental evidence suggest that $\,|\aE'(\roO)|\ll \pE'(\roO)\,$, implying that to a very good approximation
\begin{equation}
\vs = \sqrt{\pE'(\roO)}\left(1 + \frac{\aE'(\roO)}{2\,\pE'(\roO)}(\cos\theta)^2\!\right),
\label{eq:velfasea}
\end{equation}
where $\theta$ is the angle between $\bk$ and $\bno$. As regards the angular dependence of the sound speed $\vs$, \eqref{eq:velfasea} matches perfectly with the experimental findings reported in \cite{72mull} and the predictions provided by the competing second-gradient theory \cite{09virga}. In principle, the coefficient $\aE'(\roO)$ could be either positive or negative. \par
With this being said, there are several good reasons to reject the simplistic assumption \eqref{eq:T_1}. They were hinted at in Sec.\,\ref{sec:intro} and will be discussed in depth in Sec.\,\ref{sec:elas_nonhyper} and Appendix \ref{app:Ericksen}.
\subsection{A clash between anisotropy and hyperelasticity}
\label{sec:elas_nonhyper}
In a later section of the very same paper \cite{60Ericksen} where the constitutive equations leading to \eqref{eq:T_1} were proposed, after assuming a free-energy density depending only on mass density and temperature, Ericksen wrote that `[t]he form of the Clausius inequality most often used in irreversible thermodynamics [\,\dots] will lead to some restrictions on the coefficients occurring in [the above constitutive equations]' -- which he did not investigate there. In a paper he published soon after \cite{61Ericksen}, the term corresponding to the uniaxial component in \eqref{eq:T_1} was already dropped. \par
As a matter of fact, it turns out that the only way to let \eqref{eq:T_1} satisfy such restrictions is to take $\aE$ \emph{null}, which entails a spherical acoustic tensor. This is proved in Appendix \ref{app:Ericksen} in the general context of finite elasticity. That $\aE'(\roO)$ should vanish is immediately seen by linearizing \eqref{eq:T_1} around $\roO$. On account of the fact that $\roI\!=-\,\roO\!\tr\bE\,$ (with $\bE$ the infinitesimal  strain), one finds that the elastic tensor $\bChooke_{_0}\!$ equals $\roO\bAO\!\tp\bI$, with $\bAO\!$ the acoustic tensor in \eqref{eq:tensore_acustico_1} (see \cite{TP} for notations). Therefore, $\bChooke_{_0}\!$ has the major symmetry if and only if $\bAO\!$ is spherical, i.e., if and only if $\aE'(\roO)=0\,$.
\subsection{A hyperelastic transversally isotropic fluid}
\label{sec:elas_energy}
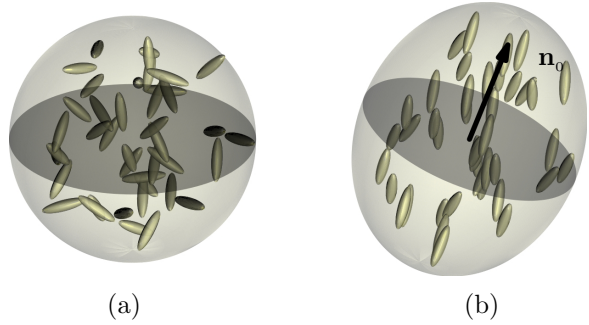
\begin{figure}[t]
\centering
\begin{picture}(250,110)(0,0)
\put(20,12){\includegraphics[width = 0.5\textwidth]{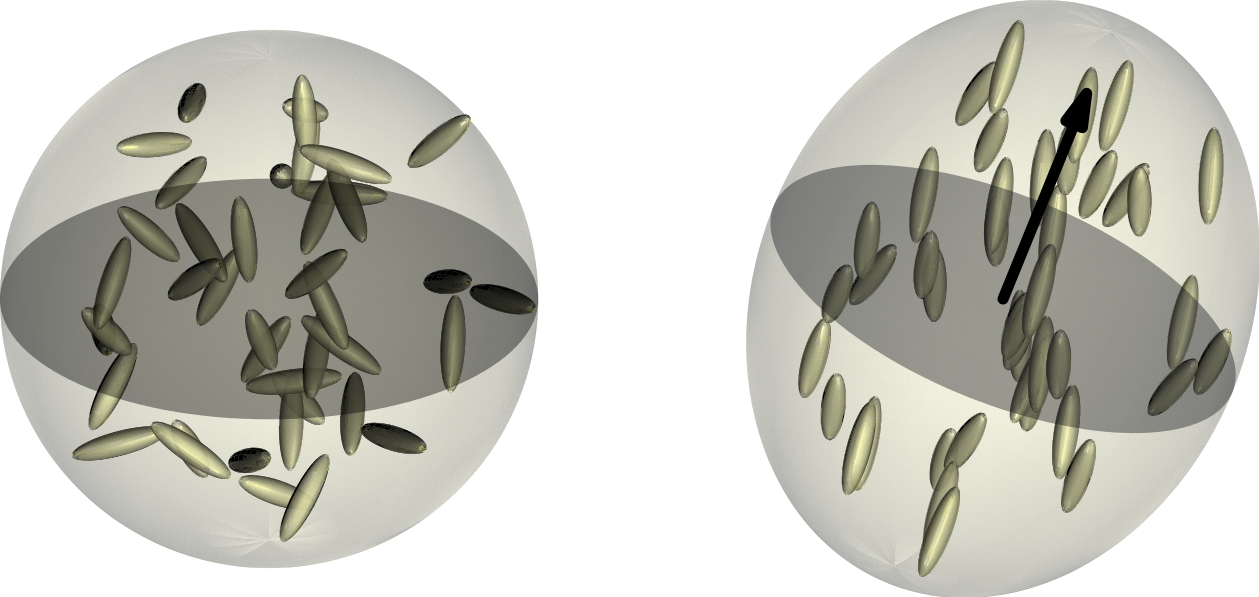}}
\put(57,-2){(a)}
\put(192,-2){(b)}
\put(220,92){$\bno$}
\end{picture}
\caption{Molecular cartoon illustrating two isoenergetic con- figurations (latent heat apart) of a small blob of liquid crystal, respectively in the isotropic (a) and in the nematic phase (b).}
\label{fig:reference}
\end{figure}

\pb{We now proceed to build-up the simplest} strain-energy density \pb{fit to describe} the anisotropic elastic response of a compressible liquid crystal in its nematic phase\pb{. To this aim} we compare its current configuration (characterized by a uniform director field) with a reference configuration in its \emph{isotropic} phase. \pb{Consider} a small LC sample whose reference shape is a spherical ball $B_{\varepsilon}$ of radius $\varepsilon$. \pb{Our strain-energy choice stems from the assumption, illustrated in Fig.\,\ref{fig:reference}, that in the nematic phase the relaxed shape of the formerly spherical ball becomes an ellipsoid of revolution sharing the same volume. Moreover, in order to represent the intrinsic \adc{anisotropy} associated with the nematic phase we introduce the \emph{anisotropic aspect ratio} $\an(\ro)$ such that the area of the cross section normal to the symmetry axis equals $\pi\varepsilon^2\!/\an(\ro)$. As a consequence, a prolate (oblate) ellipsoid is obtained whenever $\an(\
ro)$ is greater (smaller) than 1}. \pb{The} difference $\an(\ro)\!-\!1$ \pb{assumes therefore the role of} an order parameter describing the loss of spherical symmetry of the \emph{positional} pair correlation function of LC molecules due to the isotropic-to-nematic phase transition \cite{93bez,09tmbz}. \par
The above argument motivates the introduction of the following stored energy density (with respect to mass):
\begin{equation}
\sigma(\bF) =\;\sigmh(\ro,\bBo)\bydef
\sO(\ro) + \tfrac12\mu(\ro)\tr\!\big(\bCtr(\ro)^{-1}\bBo-\bI\big)\,,
\label{eq:We_prerelaxing}
\end{equation}
where $\bF$ is the deformation gradient, $\ro$ is related to it through
mass conservation: $\ro=\roO/J$, with $J\bydef\det\bF$, $\bBo\bydef\bFo\,\bFo^{\top}$\,is the \emph{left} Cauchy-Green strain tensor associated with the isochoric component of the deformation gradient $\bFo\bydef J^{-1/3}\,\bF$, and
\begin{equation}
\bCtr(\ro)\bydef\an(\ro)^2\,\bno\!\tp\bno+
\an(\ro)^{-1}\big(\,\bI-\bno\!\tp\bno\big)
\label{eq:trans_strain}
\end{equation}
is the right Cauchy-Green transformation strain from the shear-stress free configuration in the isotropic phase to the shear-stress free configuration in the nematic phase (pictorially, from (a) to (b) in Fig.\,\ref{fig:reference}), the density being $\ro$ in both of them. The nematic degrees of freedom being frozen, the director $\bno\!$ appears in \eqref{eq:trans_strain} as a parameter. \par
The first term on the right side of \eqref{eq:We_prerelaxing} is meant to account for the effects of non-volume-preserving  deformations. The second term is the simplest frame-indifferent and symmetry-allowed function of the pair $(\bFo,\bno)$ vanishing if (and only if) $\bBo=\bCtr(\ro)$. This term is formally analogous to the representation given by DeSimone and Teresi \cite{09dester} of the trace formula originally proposed by Warner, Terentjev and coworkers \cite{BTW94,WT96,WT03} to model the soft elastic response of nematic elastomers. What is specific to the present theory is the crucial dependence of the transformation strain on the density via the anisotropic aspect ratio, as established by \eqref{eq:trans_strain}. We allow also the modulus $\mu(\ro)$ to be density-dependent, but this is inessential.

\par
\RADC{As we prove in Appendix \ref{app:hyperel}, the hyperelastic theory stemming from \eqref{eq:We_prerelaxing}--\eqref{eq:trans_strain} provides the simplest extension to finite elasticity of the small-displacement theory intimated by Mullen et al.\,\cite{72mull} to make sense of their experimental results.}
\par
Assumptions \eqref{eq:We_prerelaxing}--\eqref{eq:trans_strain} entail the following prescription for the Cauchy stress (cf.~Appendix \ref{app:hyperel}):
\begin{equation}
\bT(\ro,\bBo) = -\,\ph(\ro,\bBo)\,\bI +
\ro\,\mu(\ro)\dev\!\big(\bCtr(\ro)^{-1}\bBo\big)
\label{eq:T_prerelaxing}
\end{equation}
where the product $\ro\mu(\ro)$ appears to be a shear modulus, $\dev$ is the deviatoric projector: \mbox{$\dev\bL=\bL-\tfrac13(\tr\bL)\bI$}, and
\begin{equation}
\begin{split}
\ph(\ro,\bBo) & \bydef\ro^2
\Big(\sO'(\ro)+\tfrac12\mu'(\ro)\tr\!\big(\bCtr(\ro)^{-1}\bBo-\bI\big)
\\[-.25ex]
&-\tfrac32\big(\an'(\ro)/\an(\ro)\big)\mu(\ro)
\big(\!\dev(\bno\!\tp\bno)\!\big)\!\cdot\!
\big(\bCtr(\ro)^{-1}\bBo\big)\Big).
\end{split}
\label{eq:pressure}
\end{equation}
Consistently with the result stated in Sec.\,\ref{sec:elas_nonhyper}, \eqref{eq:T_1} and \eqref{eq:T_prerelaxing} disagree unless $\mu\!=\!0$, $\aE\!=\!0$, and $\pE(\ro)\!=\!\ro^2\sO'(\ro)$. However, \eqref{eq:T_prerelaxing} does not depart too much from \eqref{eq:T_1} in the regime of interest, as we shall see in Secs.\,\ref{sec:elas_special} and \ref{sec:elas_sound}. \par
The unperturbed state of the NLC is characterized by a density equal to $\roO$ and a spherical stress tensor $\bTeq$:
\begin{equation}
\dev\bTeq\!=0\,.
\label{eq:eq_cond}
\end{equation}
\adc{If, as we assume,} $\muO\!\bydef\mu(\roO)>0\,$, \eqref{eq:T_prerelaxing} and \eqref{eq:eq_cond} imply that the left Cauchy-Green equilibrium strain $\bBoeq\!$ equals the right Cauchy-Green transformation strain:
\begin{equation}
\bBoeq\!=\bCtr(\roO)\,.
\label{eq:Boeq}
\end{equation}
Then, \eqref{eq:pressure} yields the equilibrium pressure
\begin{equation}
\pO\!\bydef\ph(\roO,\bBoeq)=\roO^2\sO'(\roO)
\label{eq:peq}
\end{equation}
and \eqref{eq:T_prerelaxing} the equilibrium stress
\begin{equation}
\bTeq\!\bydef\bT(\roO,\bBoeq)=-\,\pO\bI\,.
\label{eq:Teq}
\end{equation}
Coherently, the reference configuration is not shear-stress free in the nematic phase, unless $\aO\!\bydef\an(\roO)=1\,$:
\begin{align}
\bT(\roO,\bI)=&-\,\ph(\roO,\bI)\,\bI -\frac{(\aO\!-1)\,(1\!+\aO\!+\aO^2)}{\aO^2}\, \roO\muO\!\dev\!\left(\bno\!\tp\bno\right)
\label{eq:Tref}
\end{align}
with
\begin{align}
\ph(\roO,\bI)=\,\pO\!
+\frac{(\aO\!-1)(1+\aO\!+\aO^2)}{\aO^3}\,\roO^2\an'(\roO)\muO
+\frac{(\aO\!-1)^2\,(1+2\,\aO)}{2\,\aO^2}\,\roO^2\mu'(\roO)\,.
\label{eq:pref}
\end{align}
\subsection{A slightly compressible anisotropic fluid}
\label{sec:elas_special}
NLCs -- as all liquids -- are hard to compress: tiny density variations imply fairly large pressure changes. In addition, their faint anisotropic compressibility, however small, should be quite sensitive to small changes in density, in order to account for the angular dependence of the sound speed (cf.\,\eqref{eq:velfasea}). We wish to capture these features by specializing \eqref{eq:We_prerelaxing} accordingly. \par
First, after introducing the \emph{scaled density variation}
\begin{equation}
\xi\bydef\ro/\roO\!-1
\label{eq:sdvar}
\end{equation}
and the \emph{isotropic pressure function}
\begin{equation}
\ro\,\mapsto\piso(\ro)=\ro^2 \sO'(\ro)\,,
\label{eq:p_iso}
\end{equation}
we posit
\begin{equation}
\piso\big(\roO(1+\xi)\big) = \pO\!+ \roO\pI\xi +o(\xi)\,.
\label{eq:p_develop}
\end{equation}
The bulk modulus $\roO\pI$ is assumedly positive and much larger than the unperturbed pressure which, in turn, is much larger than the envisaged pressure perturbation:
\begin{equation}
\roO\pI\!\gg\pO\!\gg\roO\pI|\xi|\,.
\label{eq:porder}
\end{equation} \par
Next, we formalize the hypothesis that the anisotropic aspect ratio $\an(\ro)$ differs slightly from 1 \emph{uniformly} in the density $\ro$ near $\roO\!$ by positing
\begin{equation}
\an\big(\roO(1+\xi)\big)-1=\alO\!+ \alI \xi+o(\xi)
\label{eq:aaratio}
\end{equation}
and assuming
\begin{equation}
|\aO\!-1|=|\alO|\ll1\,.
\label{eq:aOsmall}
\end{equation}
In contrast to $\alO$, the \emph{sensitivity coefficient} \mbox{$\alI\!\!=\roO\an'(\roO)$} is \emph{not} required to be small, since assumption \eqref{eq:porder} ensures that $|\xi|\!\ll\!1$. This fact will play a key role in our further considerations. Henceforth, the transformation strain \eqref{eq:trans_strain} will be treated as a small perturbation of the identity. In this approximation, \eqref{eq:Boeq}, \eqref{eq:Tref} and \eqref{eq:pref} simplify respectively to
\begin{subequations}
\begin{align}
\bBoeq\!=&\;\bI+3\,\alO\!\dev\!\left(\bno\!\tp\bno\right)+o(\alO)\,,
\label{eq:Boeqa}
\\[1.75ex]
\bT(\roO,\bI)=&-\ph(\roO,\bI)\,\bI -3\,\alO
\roO\muO\!\dev\!\left(\bno\!\tp\bno\right)+o(\alO)\,,
\label{eq:Trefa} \\[1.75ex]
\ph(\roO,\bI)=&\;\pO\!
+3\,\alI\alO\roO\muO+o(\alO)\,.
\label{eq:prefa}
\end{align}
\end{subequations} \par
Finally, we posit
\begin{equation}
\mu\big(\roO(1+\xi)\big)=\muO\!+ \muI \xi+o(\xi)
\label{eq:mu_develop}
\end{equation}
and assume the shear modulus $\roO\muO\!$ to be much smaller than the bulk modulus $\roO\pI$:
\begin{equation}
\muO\!\ll\pI.
\label{eq:morder}
\end{equation}
\subsection{Angular dependence of the sound speed}
\label{sec:elas_sound}
We now look for traveling plane waves of the form
\begin{subequations}
\label{eq:Ansatz}
\begin{align}
\bu_{\eps}(\br,t) &= \eps\,\bu(\br,t) = \eps\re\!\big(\bw(\br,t)\big),
\label{eq:Ansatz_u}
\\[.25ex]
\bw(\br,t) &= \exp\!\left(i\big(\bk\!\cdot\!\br - \omega\,t\big)\!\right)\!\ba \, ,
\label{eq:Ansatz_w}
\end{align}
\end{subequations}
where $\bu_{\eps}$ is a small-amplitude displacement field, characterized by the non-dimensional smallness parameter $\eps$, the vector amplitude $\ba$, the wave vector $\bk\!=\!k\,\be$ (with $|\be|\!=\!1$ and wave number $k\!>\!0$), and the angular frequency $\omega$. All of the above quantities are real. The complex \pb{exponential form in \eqref{eq:Ansatz_w}} will turn out especially useful in Sec.\,\ref{sec:relax_sound}, where the wave vector \adc{itself will be complexified, in order to represent attenuated waves.} \par
To avoid inessential complications, we will neglect terms of order $O(\alO)$ and $o(\eps)$. Within these approximations, ansatz \eqref{eq:Ansatz} implies
\begin{subequations}
\label{eq:eps_kin}
\begin{align}
\Dv\bv&=-\,\eps\,\roO\omega^2\re(\bw)\,,
\label{eq:vdota}
\\[.5ex]
\bF & = \bI - \eps\im(\bw\tp\bk) \,,
\label{eq:eps_F}
\\[.25ex]
\xi & = \eps\im(\bw\!\cdot\!\bk)\,,
\label{eq:eps_xia}
\\[.25ex]
\bBo & = \bI-\eps\im\!\big(\!\dev(\bw\tp\bk+\bk\tp\bw)\big) \,,
\label{eq:Boa}
\\[.25ex]
\bCtr\!& = \bI+3\,\eps \,\alI\!\im(\bw\!\cdot\!\bk) \dev(\bno\!\tp\bno) \,.
\label{eq:eps_Ctra}
\end{align}
\end{subequations}
Equation \eqref{eq:eps_xia} establishes that the scaled density variation is $O(\eps)$, as expected. Substituting \eqref{eq:eps_xia}, \eqref{eq:Boa} and \eqref{eq:eps_Ctra} into \eqref{eq:T_prerelaxing}, one computes $\bT =\bTeq\!+\eps\,\bTI$, where
\begin{equation}
\begin{split}
\bTI\! & = -\roO\pI\!\im(\bw\!\cdot\!\bk)\,\bI
-\roO \muO\!\im \big\{
3\,\alI^2 (\bw\!\cdot\!\bk)\bI + \dev(\bw\tp\bk+\bk\tp\bw)
\\[.25ex]
& +3\,\alI\!\big((\dev(\bw\tp\bk))\!\cdot\!(\bno\!\tp\bno)\,\bI
+(\bw\!\cdot\!\bk) \dev(\bno\!\tp\bno)\big)\!\big\},
\end{split}
\label{eq:T1}
\end{equation}
whose divergence reads
\begin{equation}
\divr \bTI\!= - \roO\!\re\!\big(\pI\!(\bk \tp \bk) \bw+ \muO \bM \bw \big),
\label{eq:divT1}
\end{equation}
where all anisotropic information is encoded in the symmetric tensor
\begin{equation}
\bM \bydef (\bk\!\cdot\!\bk)\,\bI + (1/3 + 3\,\alI^2)(\bk \tp \bk)\,+ 3\,\alI\!\!\left((\bno\!\!\cdot\!\bk) (\bk\tp\bno\!+ \bno\tp\bk) - \tfrac{2}{3}\,\bk\tp\bk \right) .
\label{eq:tensorM_norelax}
\end{equation} \par
Substituting \eqref{eq:vdota} and \eqref{eq:divT1} into \eqref{eq:balances}$_2$ leads to the eigenvalue problem
\begin{equation}
\left(\,\pI\!(\bk \tp \bk) + \muO \bM\,\right)\ba = \omega^2\,\ba\,.
\label{eq:eigen}
\end{equation}
Assumption \eqref{eq:morder} justifies introducing another small parameter -- reminiscent of the small ratio $|\aE'(\roO)|/\pE'(\roO)$ in Sec.\,\ref{sec:elas_non-hyper} -- namely
\begin{equation}
\eta\bydef\muO/\pI.
\label{eq:eta}
\end{equation}
Expanding the wavelength $\lambda=2\pi/k$ and the wave amplitude vector $\ba$ in terms of $\eta$, we get
\begin{subequations}
\begin{align}
\lambda_{\eta}\!& = {\elO + \eta\,\elI\!+ o(\eta)}\,,
\\
\ba_{_\eta}\!\!& = \baO\!+ \eta\,\baI\!+ o(\eta)\,.
\end{align}
\label{eq:lam&a_exp}
\end{subequations}
The $O(1)$ solution to \eqref{eq:eigen} yields a longitudinal wave vector $\baO\!$ and an isotropic sound speed $\vO$, as expected:
\begin{equation}
\baO\!=\AO\be\;\;\,\textrm{with}\;\AO\!>0\,,\qquad
\vO\!=\frac{\omega}{2\pi}\,\elO\!=\sqrt{\,\pI}\,.
\label{eq:O(1)}
\end{equation}
The $O(\eta)$ problem reads
\begin{equation}
-2(\elI/\elO)\AO\be+(\elO/2\pi)^2\AO\bMo\be=
( \bI - \be \tp \be )\,\baI,
\label{eq:order1_eta}
\end{equation}
with
\begin{align}
(\elO/2\pi)^2\,\bMo\be = \big(4/3-2\,\alI\!+3\,\alI^2
+3\,\alI\!(\cos\theta)^2\big)\be + 3\,\alI\!(\cos\theta)\bno\,,
\label{eq:order0_M}
\end{align}
where $\theta$ is the angle between $\be$ and $\bno$. The solubility condition of \eqref{eq:order1_eta} yields the $O(\eta)$ anisotropic correction to the speed of sound:
\begin{equation}
\vs/\vO=1+
\eta\big(2/3-\alI\!+3\,\alI^2/2 + 3\,\alI (\cos \theta)^2 \big)\,.
\label{eq:speed_elastic}
\end{equation}
Then, solving \eqref{eq:order1_eta} (in the subspace orthogonal to $\be$) delivers the $O(\eta)$ correction to the wave amplitude vector:
\begin{equation}
\be\!\times\!\baI\!= 3\,\alI\!\AO\!(\cos\theta)\,\be\!\times\!\bno .
\label{eq:order1_a}
\end{equation}
Solution \eqref{eq:order1_a} exhibits the existence of an $O(\eta)$ transversely polarized component in the plane spanned by $\bno$ and $\be$, whose amplitude is maximal when $\theta=\pi/4\,$:
\begin{equation}
\baI\!= \tfrac32\,\alI\AO\!(\sin2\,\theta)\bt\,,
\label{eq:trans_a}
\end{equation}
with $\bt$ the unit vector orthogonal to $\be$ in $\spn\{\bno,\be\}$ such that $\bt\!\cdot\!\bno\!>\!0\,$.
The wave amplitude vector $\ba$ is slightly tilted towards the nematic director if $\alI\!>0\,$, away from it if $\alI\!<0\,$. \par
As far as the angular dependence of the sound speed is concerned, \eqref{eq:speed_elastic} is completely satisfactory. However, the underlying model is obviously unable to account for the important frequency dependence and the related attenuation sistematically observed in experiments \cite{72mull,70lord,71lieb,71kemp}. This issue will be attacked in Sec.\,\ref{sec:relax}. \par
A more accurate calculation would add insignificant terms of order $O(\alO)$ to the coefficient of $\eta$ in \eqref{eq:speed_elastic}, changing it into
\begin{equation}
2/3-\alI\!+3\,\alI^2/2+O(\alO) + (3+O(\alO)\!)\,\alI (\cos \theta)^2.
\label{eq:speed_elastic_acc}
\end{equation}
We conclude that, while the precise value of the anisotropic aspect ratio is immaterial (as long as it is near to 1), what really matters is the value of its derivative. In particular, the speed of sound is maximal or minimal when the direction of propagation is along the nematic director, depending on whether the anisotropic aspect ratio grows or decreases with increasing density. The experimental evidence reported in \cite{72mull,70lord} points to a positive value of $\alI$.
\section{Nematic relaxation}
\label{sec:relax}
The hyperelastic theory set up in Sec.\,\ref{sec:elas_energy} misses the effects of molecular rearrangements which, while not affecting the macroscopic deformation of the NLC, do make the ensuing shear stress relax to zero whenever its macroscopic evolution is sufficiently slow. Roughly speaking, these microscopic rearrangements drive the liquid to lower energy states. A convenient caricature of their macroscopic effects is obtained by postulating that the relaxed configuration evolves on a characteristic time scale, steering towards a moving target, the actual configuration. This establishes a competition between the relaxation and the loading dynamics, which may explain the frequency dependence of the effective elastic moduli. Interestingly, a similar idea was already outlined in the last section of the seventh volume of the Course of Theoretical Physics by Landau and Lifshitz \cite{59L&L}. \par
\begin{figure}[h]
\centering
\begin{picture}(250,135)(0,0)
\put(20,-4){\includegraphics[width=0.5\textwidth]{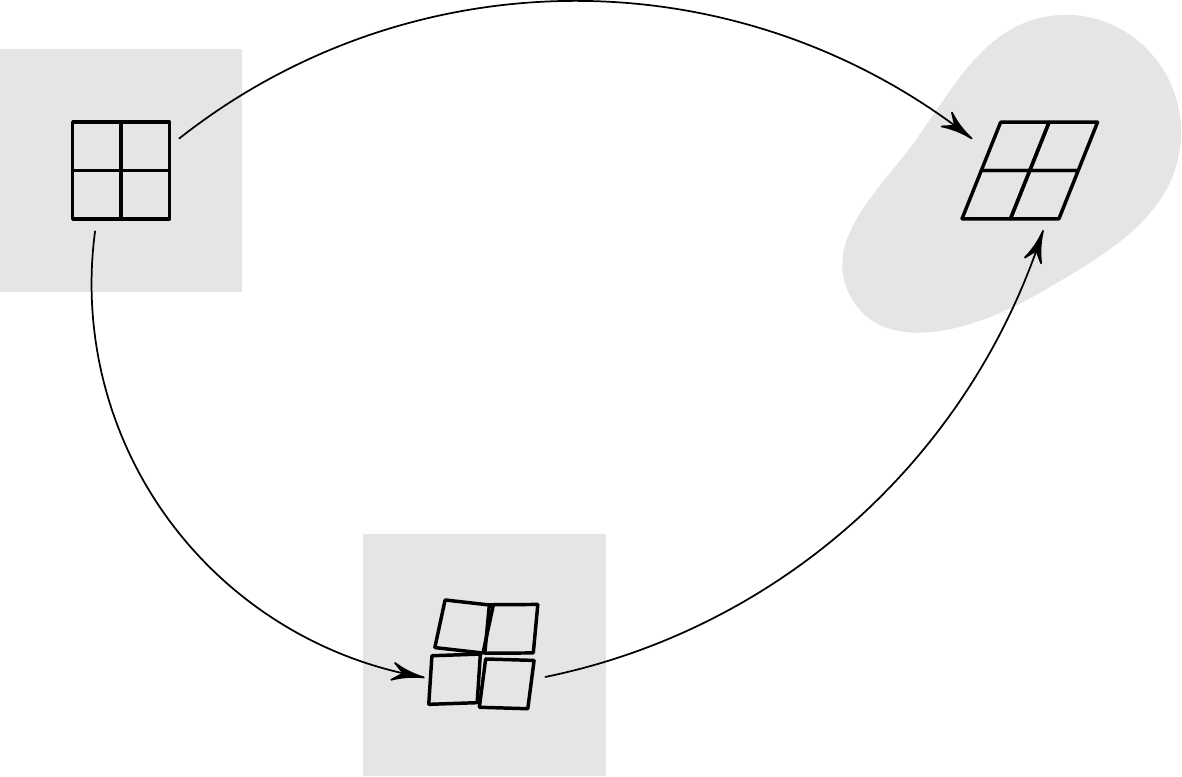}}
\put(22.5,122){\emph{reference}}
\put(95,0){\emph{relaxed}}
\put(203,122){\emph{actual}}
\put(124,128){$\bF$}
\put(42,31){$\bG$}
\put(178,31){$\bFe\!\bydef\bF\bG^{-1}$}
\end{picture}
\caption{Instantaneous decomposition of the deformation gradient $\bF$ into a relaxing and an effective component. Disarrangements in the lower cartoon are meant to illustrate the fact that, contrary to $\bF$, the relaxing tensor field $\bG$ and the effective deformation field $\bFe$ need not be gradients.}
\label{fig:patatoidi}
\end{figure}
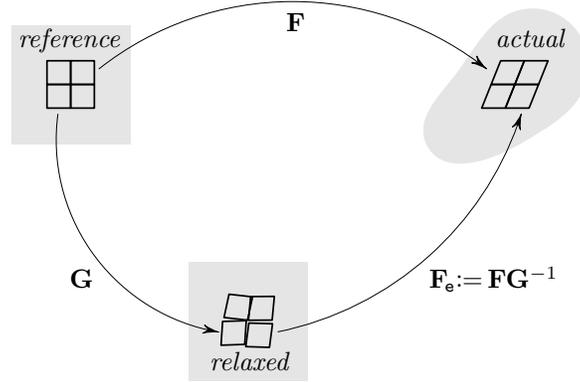

With these motivations, we \adc{split the deformation gradient $\bF$ into the product of a \emph{relaxing deformation} $\bG$ and an \emph{effective deformation} $\bFe$ (cf.~Fig.\,\ref{fig:patatoidi}):}
\begin{equation}
\adc{
\label{eq:Fe}
\bF=\bFe\bG\,,
}
\end{equation}
and let the stored energy depend only on the latter (as described below). The relaxing deformation $\bG$ is a time-dependent tensor field,  taking values in the proper special linear group $\SLp$ (the Lie group of double tensors whose determinant equals~$+1$). This modeling choice is prompted by the experimental fact that no molecular rearrangement can accommodate density variations and, as a consequence, only the deviatoric component of the stress may relax to zero. Since the relaxed configuration is defined only elementwise \cite{02dicqui}, in \eqref{eq:Fe} the only gradient is $\bF$: the tensor field $\bG$ need \emph{not} be a gradient, and the effective deformation $\bFe$ has to compensate its integrability defect. \par
After introducing the \emph{inverse relaxing} right Cauchy-Green strain
\begin{equation}
\label{eq:H}
\bH\bydef\!\left(\bG^{\!\top}\bG\right)^{-1}
\end{equation}
\adc{and the \emph{isochoric} left Cauchy-Green strain tensor}
\begin{equation}
\label{eq:delbe}
\bBoe\bydef\,\bFoe\bFoe^{\top}\!=\bFo\,\bH\,\bFo^{\top},
\end{equation}
we simply substitute $\bBo$ with $\bBoe$ in \eqref{eq:We_prerelaxing}, obtaining the stored energy density
\begin{align}
\sigmr(\bF,\bH)\bydef&\,\sigmh(\ro,\bBoe) = \;\sO(\ro) +\tfrac12\mu(\ro)\tr\!\big(\bCtr(\ro)^{-1}\bBoe-\bI\big),
\label{eq:We_relaxing}
\end{align}
where $\ro=\!\roO/J\!=\!\roO/(\det\bFe)$. Since the relaxing deformation $\bG$ enters \eqref{eq:We_relaxing} only through the corresponding $\bH$, we are naturally led to introduce the manifold $\Rlx$ of relaxing strains as the intersection of the proper special linear group $\SLp$ and the space of symmetric double tensors $\Sym$:
\begin{equation}
\Rlx\,\bydef\,\SLp\!\cap\Sym\,.
\label{eq:Rlx}
\end{equation}
 \par
The Cauchy stress is still given by \eqref{eq:T_prerelaxing}, provided that $\bBo$ is substituted by $\bBoe$:
\begin{equation}
\label{eq:T_relaxing}
\bT(\ro,\bBoe) = -\,\ph(\ro,\bBoe)\,\bI +
\ro\,\mu(\ro)\dev\!\big(\bCtr(\ro)^{-1}\bBoe\big)
\end{equation}
Of course, balance equations \eqref{eq:balances}, \adc{supplemented} with the constitutive prescription \eqref{eq:T_relaxing}, need to be complemented by an evolution equation along $\Rlx$ governing the relaxational degrees of freedom.

\subsection{Relaxation dynamics}
\label{sec:relax_eq}
Inspired by ideas presented by Rajagopal and Srinivasa \cite{98rajaII} and by DiCarlo and coworkers \cite{02dicqui,05adc,07desdicter}, we hypothesize a viscous-like dynamics for the inverse relaxing strain, described by a steepest-descent equation on the relaxing strain manifold $\Rlx$:
\begin{equation}
\gamma\,\dot{\bH}=-\,\roO\PH\frac{\partial\sigmr}{\partial\bH}\;.
\label{eq:steep}
\end{equation}
The scalar coefficient $\gamma\!>\!0$ is a viscosity modulus, and $\PH\!$ is the orthogonal projector from the space of double tensors onto the subspace $\THR$ tangent to $\Rlx$ at $\bH$:
\begin{equation}
\PH\!=\sym - \frac{\bH^{-1}\tp\,\bH^{-1}}{\|\bH^{-1}\|^2}\,,
\label{eq:P}
\end{equation}
with $\sym$ the orthogonal projector onto $\Sym$ (see \cite{TP} for notations). Equation \eqref{eq:P} is an easy consequence of the formula for the derivative of the determinant:
\begin{equation}
\det(\bH\!+\!\eps\,\bL)=(\det\bH)(1+\eps\,\bH^{-\tsp}\!\cdot\!\bL)+o(\eps)
\label{eq:Ddet}
\end{equation}
($\,=1+\eps\,\bH^{-1}\!\cdot\!\bL+o(\eps)$, since $\bH\!\in\!\Rlx\,$), implying that the unit normal to $\THR$ is $\bH^{-1}/\|\bH^{-1}\|$. \par
The evolution equation \eqref{eq:steep} is consistent with the dissipation principle establishing that the power dissipated -- defined as the difference between the power expended and the time derivative of the free energy -- should be non-negative, for all body-parts, at all times. This condition localizes into:
\begin{equation}
\label{eq:diss}
\bS\!\cdot\!\dot{\bF}-\roO\dot{\sigmr}\ge0\,,
\end{equation}
with $\bS$ the Piola stress. Since $\bS\!=\!\roO\!\left(\partial\sigmr/\partial\bF\right)$ (cf.~Appendix \ref{app:hyperel}), \eqref{eq:diss} reduces to
\begin{equation}
\label{eq:rediss}
\roO\frac{\partial\sigmr}{\partial\bH}\!\cdot\!\dot{\bH}\le0\,.
\end{equation}
Hypothesis \eqref{eq:steep} -- sometimes justified by the heuristic criterion of maximum rate of dissipation \cite{98rajaII} -- satisfies requirement \eqref{eq:rediss} in the simplest possible way. \par
On account of \eqref{eq:We_relaxing} and \eqref{eq:delbe}, we obtain explicitly
\begin{equation}
\frac{\partial\sigmr}{\partial\bH}=
\tfrac12\mu(\ro)\bFo^{\tsp}\bCtr(\ro)^{-1}\bFo\in\Sym
\label{eq:diffexpl}
\end{equation}
which, substituted in \eqref{eq:steep}, yields
\begin{equation}
\frac{2\,\gamma}{\roO\mu(\ro)} \,\dot{\bH} -
\frac{\bCtr(\ro)^{-1}\!\cdot\!\big(\bBo\,\bBoe^{-1}\bBo\big)}
{\|\bH^{-1}\|^2}\,\bH^{-1}
= -\,\bFo^{\tsp}\bCtr(\ro)^{-1}\bFo \, .
\label{eq:steep_explicit}
\end{equation}
\subsection{Frequency-dependent anisotropic sound speed}
\label{sec:relax_sound}
\adc{We return to the problem studied in Section~\ref{sec:elas_sound}, looking now for \emph{attenuated} plane waves. We keep \eqref{eq:Ansatz_u} as is, but we change \eqref{eq:Ansatz_w} into}
\begin{equation}
\label{eq:ansatz_wr}
\bw(\br,t) = \exp\!\big(i\big(\bkc\!\cdot\!\br - \omega\,t\big)\!\big)\ba\,,
\end{equation}
\adc{where}
\begin{equation}
\bkc\bydef\bk + i\bl\quad(\bk,\!\bl\;\textrm{real})
\end{equation}
\adc{is a \emph{complex} wave vector, whose real part $\bk$ parameterizes the propagation direction and the wavelength, exactly as in Section~\ref{sec:elas_sound}, while its imaginary part $\bl$ determines how fast the  longitudinal and transverse components of the wave get damped. This viscous effect is the upshot of the relaxation dynamics introduced in Sec.\,\ref{sec:relax_eq}.} \par
As in Sec.\,\ref{sec:elas_sound}, we shall neglect terms of order $O(\alO)$ and $o(\eps)$. The deformation gradient $\bF$, the scaled density variation $\xi$ and the transformation strain $\bCtr(\ro)$ are still represented by \eqref{eq:eps_F}, \eqref{eq:eps_xia} and \eqref{eq:eps_Ctra}, respectively, provided that $\bk$ is substituted by $\bkc$ and \eqref{eq:Ansatz_w} by \eqref{eq:ansatz_wr}.

At equilibrium, the inverse relaxing strain $\bH$ equals the identity $\bI$. After positing
\begin{equation}
\label{eq:H1def}
\bH\!=\!\bI+\eps\,\bHI\quad\textrm{with }\,\bHI\!\in\TIR
\end{equation}
(i.e., symmetric and traceless), we linearize  \eqref{eq:steep_explicit} accordingly, obtaining
\begin{equation}
\label{eq:steep_lin}
\tau\,\bHId\!+ \bHI = \im(\bK\!+\!3\alI\!\bN)\,,
\end{equation}
after introducing the \emph{relaxation time}
\begin{equation}
\label{eq:relaxtime}
\tau\bydef\,\frac{2\,\gamma}{\roO\muO}
\end{equation}
and positing
\begin{subequations}
\label{eq:K&N}
\begin{align}
\bK&\bydef\dev(\bw\tp\bkc+\bkc\tp\bw)\,,
\label{eq:K}
\\[.125ex]
\bN&\bydef\,(\bw\!\cdot\!\bkc)\dev(\bno\!\tp\bno)\,.
\label{eq:N}
\end{align}
\end{subequations}
Modulo an exponentially decaying transient, \eqref{eq:steep_lin} is solved by
\begin{equation}
\label{eq:H1sol}
\bHI\!= \im\!\left((1\!-i\omega\tau)^{-1}(\bK\!+\!3\alI\!\bN)\right).
\end{equation}
Consequently, from \eqref{eq:H1def}, \eqref{eq:delbe} and \eqref{eq:T_relaxing} we get the effective isochoric strain
\begin{equation}
\bBoe = \bI + \eps\big(\bHI\!-\im(\bK)\big)
= \bI + \eps\im\!\left((1\!-i\omega\tau)^{-1}
(i\omega\tau\bK+3\alI\!\bN)\right)
\label{eq:Be_rel}
\end{equation}
and the $O(\eps)$ increment of the Cauchy stress
\begin{equation}
\begin{split}
\bTI\! & = -\roO\pI\!\im(\bw\!\cdot\!\bkc)\,\bI +\roO \muO\!\im\!\big\{i\omega\tau\,(1\!-i\omega\tau)^{-1} \big(\,3\,\alI^2 (\bw\!\cdot\!\bkc)\bI + \bK \\[.25ex]
& +3\,\alI\!\big((\dev(\bw\tp\bkc))\!\cdot\!(\bno\!\tp\bno)\,\bI
\,+\bN\big)\!\big\},
\end{split}
\label{eq:T1_relax}
\end{equation}
whose divergence reads
\begin{equation}
\divr \bTI\!= - \roO\!\re\!\left(\!\pI\!(\bkc \tp \bkc) \bw
-\frac{i\omega \tau}{1\!-i\omega \tau}\,\muO \bM \bw\!\right) ,
\label{eq:divT1_relax}
\end{equation}
where the symmetric tensor $\bM$ is still defined as in \eqref{eq:tensorM_norelax}, provided that $\bk$ is substituted by $\bkc$. In the asymptotic limit $\,\omega\tau\to\infty\,$, the solid-like elastic response analyzed in Sec.\,\ref{sec:elas_sound} is recovered:
$\bHI\,$tends to zero, and \eqref{eq:Be_rel}, \eqref{eq:T1_relax} and \eqref{eq:divT1_relax} tend to \eqref{eq:Boa}, \eqref{eq:T1} and \eqref{eq:divT1}, respectively. \par
\begin{figure}[h]
\centering
\begin{picture}(250,140)(0,0)
\put(30,0){\includegraphics[width = 0.4\textwidth]{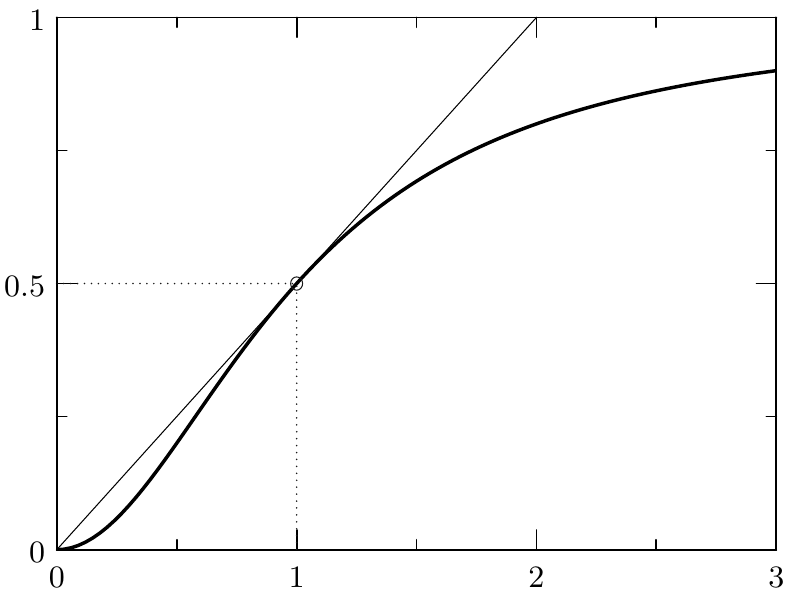}}
\put(177,2){$x$}
\put(155,97){$f$}
\end{picture}
\caption{Plot of the modulating function \eqref{eq:modulating} in the range $[0,3]$. The tangent to its graph at $(1,f(1)\!=\!1/2)$ is highlighted. Note that it hits the origin.}
\label{fig:omegatau}
\end{figure}

In complete analogy with the procedure in Sec.\,\ref{sec:elas_sound}, we substitute \eqref{eq:vdota} and \eqref{eq:divT1_relax} into \eqref{eq:balances}$_2$ and obtain the complex eigenvalue problem
\begin{equation}
\Big(\bkc \tp \bkc - \frac{i\omega \tau}{1\!-i\omega \tau}\,\eta\,\bM\Big)\ba = (\omega^2\!/\pI)\,\ba\, ,
\label{eq:eigen_relax}
\end{equation}
where $\eta$ is the small \Radc{non-dimensional} parameter introduced in \eqref{eq:eta}. After supplementing expansions \eqref{eq:lam&a_exp} with
\begin{equation}
\bl_{_\eta}\!\! = \blO\!+ \eta\blI\!+ o(\eta)\,,
\label{eq:l_exp}
\end{equation}
we perform a perturbation analysis of \eqref{eq:eigen_relax}, split  into its real and imaginary parts. At order $O(1)$, we obtain $\blO\!=\!0$ and recover the same wave vector and isotropic sound speed given in \eqref{eq:O(1)}. The real $O(\eta)$ problem reads
\begin{equation}
2\,(\elI/\elO)\,\AO\be -
\Big(\frac{\elO}{2\pi}\Big)^{\!2}\AO\!
\re\!\left(\frac{i\omega\tau}{1\!-i\omega\tau}\right)\!\bMo\be\,,
= (\bI - \be \tp \be )\,\baI,
\label{eq:order1_eta_relax}
\end{equation}
with $\bMo\be$ given by \eqref{eq:order0_M}. The solubility condition of \eqref{eq:order1_eta_relax} yields the $O(\eta)$ frequency-dependent, anisotropic correction to the speed of sound:
\begin{equation}
\frac{\vs}{\,\vO}=1+
\eta\,f(\omega\tau)
\Big(\tfrac23-\alI\!+\tfrac32\,\alI^2 + 3\,\alI (\cos \theta)^2 \Big),
\label{eq:speed_relax}
\end{equation}
with
\begin{equation}
f(x)\bydef\frac{x^2}{1+x^2}\quad(x\ge0)\,.
\label{eq:modulating}
\end{equation}
The modulating function $f$ behaves like $x\mapsto x^2$ for $x\to0$ and like $x\mapsto 1\!-\!(1/x)^2$ for $x\to\infty\,$. More interestingly, it is well approximated by the \emph{linear} function $x\mapsto x/2$ in a rather large neighborhood of $x\!=\!1$ (cf. Fig.\,\ref{fig:omegatau}).
As we shall discuss in depth in Sec.\,\ref{sec:discussion}, this appears to be the regime relevant to the experiments motivating our modeling effort \cite{72mull,70lord}. The same modulation affects also the $O(\eta)$ correction to the wave amplitude vector:
\begin{equation}
\label{eq:order1_a_relax}
\baI\!=\tfrac32\,\alI\AO f(\omega\tau)(\sin2\,\theta)\bt\,,
\end{equation}
where $\bt$ is the unit vector normal to $\be$ introduced in \eqref{eq:trans_a}. Finally, solving the imaginary $O(\eta)$ problem
\begin{equation}
(\bI+\be\tp\be)\!\blI\!= \frac{\elO}{2\pi}
\im\!\left(\!\frac{i\omega\tau}{1\!- i\omega\tau}\!\right)\!\bMo\be\, ,
\end{equation}
delivers the attenuation vector
\begin{align}
\bl & =\eta\im\!\left(\!\frac{i\omega\tau}{1\!- i\omega\tau}\!\right)\!
\frac{\elO}{2\pi}\big(\bI -\tfrac12 \be \tp \be\big) \bMo\be
\notag \\[.5ex]
& = \eta\,\frac{\omega\tau}{1\!+(\omega\tau)^2}\,\frac{2\pi}{\elO}
\Big( \big(\tfrac23 - \alI\!+\tfrac32\,\alI^2+3\,\alI\!(\cos\theta)^2\big)\be +\tfrac32\,\alI\!(\sin2\theta)\bt\Big).
\label{eq:att_vec}
\end{align}
It is noteworthy that the longitudinal and the transverse component of the attenuation vector depend on the angle $\theta$ exactly as the sound speed (cf.\,\eqref{eq:speed_relax}) and the amplitude of the transverse wave (cf.\,\eqref{eq:order1_a_relax}), respectively. Since Lord and Labes \cite{70lord} give attenuation data in decibels per unit flight time rather than distance traveled, it is apposite to transform \eqref{eq:att_vec} accordingly, by multiplying the attenuation vector by the isotropic sound speed \eqref{eq:O(1)}$_2$:
\begin{align}
\vO\!\bl\; = (\eta/\tau)f(\omega\tau)
\Big( \big(\tfrac23 - \alI\!+\tfrac32\,\alI^2+3\,\alI\!(\cos\theta)^2\big)\be +\tfrac32\,\alI\!\adc{(\sin 2\theta)}\bt\Big),
\label{eq:att_freq}
\end{align}
with $f$ the modulating function \eqref{eq:modulating}. These and the previous results will be interpreted and compared with the cited experimental data in Sec.\,\ref{sec:discussion}.
\section{Discussion}
\label{sec:discussion}
After offering scattered comments on the qualitative agreement of our results with the experimental evidence published in the early seventies \cite{72mull,70lord,71lieb,71kemp}, we now proceed to a quantitative comparison between the predictions of our theory with the data by Lord and Labes \cite{70lord} and Mullen et al. \cite{72mull}. Both groups experimented on the same LC molecule, namely, N-(4-methoxybenzylidene)-4-butylaniline (MBBA) \cite{PAA}. Since their data have been used also by De~Matteis and Virga \cite{11virga} to estimate the phenomenological parameters of their second-gradient theory, this will facilitate the comparison between our theory and theirs. \par
Our retrodictions, encapsulated in \eqref{eq:speed_relax} and \eqref{eq:att_freq}, depend on a handful of parameters: the unperturbed mass density $\roO$, the bulk modulus $\roO\pI$, the shear modulus $\roO\muO$, the anisotropic sensitivity coefficient $\alI$, and the relaxation time $\tau$. The mass density is a standard material property: we take
\begin{equation}
\label{eq:val_density}
\roO\!=10^3\,\kg/\m^3.
\end{equation} \par

Lord and Labes \cite{70lord} report that `[b]ackground attenuation varied from 2.18\,dB/$\mu$s at 6\,MHz to 0.37\,dB/$\mu$s at 2\,MHz.' This information, plugged into \eqref{eq:att_freq}, translates into the equation
\begin{equation}
\label{eq:L&L_ratio}
\frac{f(12\cdot\!10^6\,\pi\,\tau)}{f(4\cdot\!10^6\,\pi\,\tau)}=\frac{2.18}{0.37}
\end{equation}
(with $f$ the modulating function \eqref{eq:modulating} and $\tau$ in seconds), whose solution yields the relaxation time
\begin{equation}
\label{eq:val_relax}
\tau=2.11\!\cdot\!10^{-8}\,\mathrm{s}\,.
\end{equation}
Remarkably, this value coincides with the one ($2\cdot\!10^{-8}\,\mathrm{s}$) estimated on different grounds by Mullen et al.\,\cite{72mull}. Differently from them, we do not assume as a hypothesis that $\omega\tau\!\approx\!1$, but find as a result that all of the experiments reported in \cite{72mull,70lord} fall in the interval
\begin{equation}
\label{eq:interval}
0.265\le\omega\tau\le1.86\,.
\end{equation}
We consider therefore our estimate \eqref{eq:val_relax} decidedly robust. \par
We now let $\alI\!\!=0$ in \eqref{eq:att_freq}, presuming that this somehow mimics the `background attenuation' measured by Lord and Labes. This is a tentative hypothesis, since our model ignores all dissipation mechanisms different from the relaxation dynamics introduced in Sec.\,\ref{sec:relax_eq}. In particular, it lacks the contribution of the Leslie-Ericksen viscous stress tensor, extended to compressible LCs \cite{75monroe} -- which, contrariwise, is the only source of dissipation in the competing second-gradient theory \cite{09virga,11virga}. At any rate, the set of two equations stemming from the above hypothesis is solved by the value of $\tau$ in \eqref{eq:val_relax} and the (reasonably small) dimensionless ratio
\begin{equation}
\label{eq:val_ratio}
\eta=2.05\cdot\!10^{-2}.
\end{equation} \par
To estimate the key parameter of our model, namely, the anisotropic sensitivity coefficient $\alI$, we make recourse to the data from Fig.\,2 of \cite{72mull} establishing that the experimental angular dependence of sound velocity -- defined as the difference between the velocity in the direction at an angle $\theta$ with respect to $\bno\,$and the velocity in the direction normal to it, divided by the absolute velocity -- at $21\,\Cels$ and 10\,MHz is nicely interpolated by $1.25\cdot\!10^{-3}(\cos\theta)^2$. As for the absolute sound velocity, we may use the value $1.54\cdot\!10^3\,\mathrm{m/s}$ measured by Mullen et al.\,\cite{72mull} for 2\,MHz at $22\,\Cels$, since scaling it to 10\,MHz according to \eqref{eq:speed_relax} affects it far beyond the third significant digit. By plugging the above data into \eqref{eq:speed_relax}, we obtain the estimate
\begin{equation}
\label{eq:val_al1}
\alI\!=3.19\cdot\!10^{-2}.
\end{equation} \par
Then, using \eqref{eq:speed_relax} and \eqref{eq:O(1)}$_2$ yields
\begin{equation}
\label{eq:val_p1}
\pI\!=2.37\cdot\!10^6\,\mathrm{m^2\!/s^2},
\end{equation}
which, combined with \eqref{eq:val_density}, \eqref{eq:val_ratio} and \eqref{eq:eta}, delivers the following estimates for the bulk and shear moduli:
\begin{subequations}
\label{eq:val_moduli}
\begin{align}
\roO\pI\!&=2.37\,\mathrm{GPa}\,,
\label{eq:val_bulkm}
\\
\roO\muO\!&=48.6\,\mathrm{MPa}\,.
\label{eq:val_bulks}
\end{align}
\end{subequations} \par
\begin{figure}[t]
\centering
\begin{picture}(250,160)(0,0)
\put(18,-68){\includegraphics[width = 0.5\textwidth]{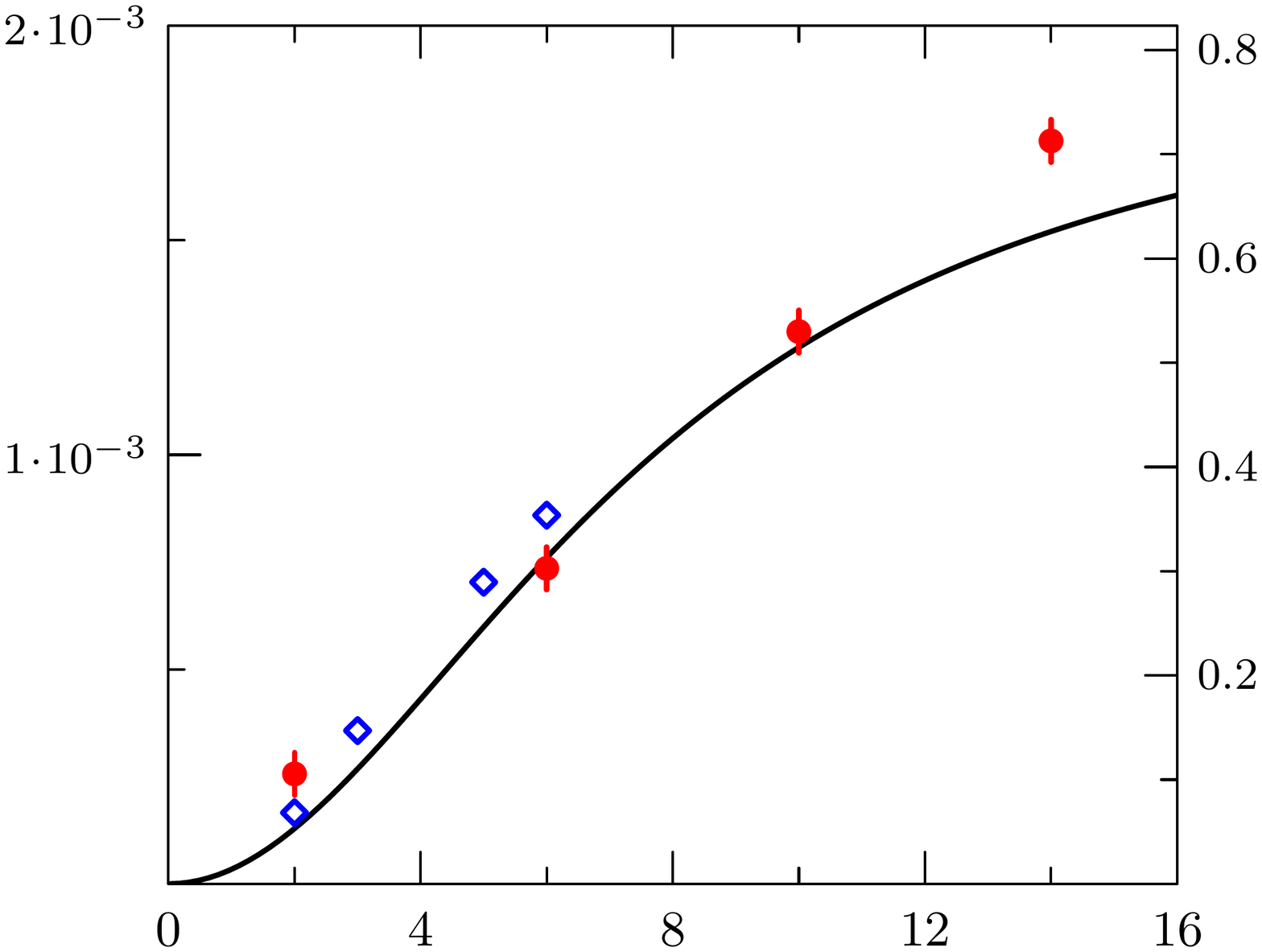}}
\put(144,0){MHz}
\put(227,140){dB/$\mu$s}
\end{picture}
\caption{Frequency dependence of the velocity anisotropy (left $y$-axis) and the attenuation anisotropy, defined as the difference between the attenuation along $\bno\,$and the attenuation in the direction normal to it (right $y$-axis). Circles with error bars represent experimental values of the velocity anisotropy taken from Fig.\,3 of \cite{72mull} (at a temperature of $27\,\Cels$). Diamonds reproduce the attenuation data from Fig.\,2 of \cite{70lord}. The full line is the retrodiction of our theory for both velocity and attenuation anisotropy.}
\label{fig:confronto}
\end{figure}

A direct quantitative comparison between our results and experimental data from \cite{72mull,70lord} is provided in Fig.\,\ref{fig:confronto}. Their less-than-perfect agreement is more than satisfactory, on account of the fact that our theoretical curve, far from being fitted to the eight experimental points in the figure, depends on only five parameters -- two of which standard -- that have been identified on the basis of experimental information from the same sources \cite{72mull,70lord}, but \emph{independent} of the data gathered in Fig.\,\ref{fig:confronto}. \par
Clearly, the weak point of the above estimation procedure lies in the identification of the background attenuation leading to \eqref{eq:val_ratio}. This points to the need of extending the present model to include the Leslie-Ericksen viscosities and the interest of studying how they interact with the anisotropic compressibility peculiar to our theory. This would likely introduce at least one more relaxation time. \RADC{But even in its present state, our model provides the first quantitatively based explanation of the structural relaxation process hypothesized long ago in \cite{72mull}.} \par
Also the constraint on the nematic director should be lifted, adding an Oseen-Frank term to the stored energy density, as sketched in Appendix~\ref{app:hyperel}. \adc{However, one of us has already proved \cite{XXturzi} that director oscillations do not affect significantly the present results.} \par
\pb{\Radc{Finally, we point out that the present theory could} be readily generalized to encompass temperature-dependent effects, including the critical behavior close to the isotropic-to-nematic phase transition. To this end, one should first replace the nematic director with de Gennes' nematic order tensor $\bQ$. This may be done simply by substituting the \adc{$\bno\!\tp\bno$\,terms in \eqref{eq:trans_strain} -- or the $\bn\tp\bn$ terms in \eqref{eqapp:trans_strain} --} with $\bQ+\tfrac13\bI$, and adding the usual Landau-de Gennes thermodynamic potential to the free energy. This generalization, however, has to be handled with care, since the coefficients of the thermodynamic potential depend on density in a critical way. Indeed, as already establised by early Onsager's theories \cite{Ons}, even tiny density changes can strongly affect the critical behavior of NCLs.}
\section*{Acknowledgements}
\adc{We} wish to thank Luciano Teresi for enlightening discussions on various topics touched upon in this paper. \pb{Financial support from the Italian Ministry of University and Research through the Grant No.\,200959L72B\underline{\;\;}004 `Mathematics and Mechanics of Biological Assemblies and Soft Tissues' is gratefully acknowledged.}


\begin{thebibliography}{10}
\bibitem{Note} In the experiments reported in \cite{72mull} the temperature was stabilized to within $\pm0.1\,\Cels$ or less. Lord and Labes \cite{70lord} experimented at room temperature. What was actually measured by both groups was the angle between the propagation direction and the direction of the uniform magnetic field used to align the molecules. Lord and Labes \cite{70lord} applied fields up to 12\,kOe, observing no change in the attenuation beyond~1\,kOe. Mullen, L\"{u}thi and Stephen  \cite{72mull} reported no field dependence between 0.5 and 5\,kOe. Also the nematoacoustic analysis by Virga \cite{09virga} is based on the hypothesis of a uniform and immobile director field. Lifting this assumption, as done by De~Matteis and Virga \cite{11virga}, does not seem to make much of a difference -- at least within the second-gradient theory developed in \cite{09virga}.
\bibitem{TP} We define the tensor product between two vectors $\ba,\bb$ as follows: $(\ba\tp\bb)\,\bu=(\bb\!\cdot\!\bu)\,\ba$ for all $\bu$ (i.e., $(\ba\tp\bb)_{ij}\!={\rm a}_{i}{\rm b}_{j}$ in any orthonormal basis). The tensor product between double tensors is defined analogously, so that $(\bA\otimes\bI)\,\bE=(\bI\!\cdot\!\bE)\,\bA\!=\!(\tr\bE)\,\bA$, $\bI$ being the identity on vectors. The inner product between double tensors is defined as $\bM\!\cdot\!\bL=\tr\,(\bM^{\top}\bL)$. The trace $\tr$ is the unique linear form on double tensors such that $\tr(\ba\tp\bb)=\ba\!\cdot\!\bb={\rm a}_{i}{\rm b}_{i}$ (sum over $i$). We~use few fourth-order tensors, denoted by blackboard bold symbols (apart from the projectors \Radc{$\dev$, $\sym$, and $\skw$}).
\end{thebibliography}
%
%
\appendix
\numberwithin{equation}{section}
\section{Ericksen's elastic fluid is hyperelastic only if isotropic}
\label{app:Ericksen}
Ericksen's constitutive assumption \eqref{eq:T_1} for the Cauchy stress
translates into the following prescription for the Piola stress (cf.~Appendix \ref{app:hyperel}):
\begin{equation}
\bS=\phi(J)\, \bF^{-\tsp}+ \psi(J) \,\bno\!\tp\left(\bF^{-1}\bno\right),
\label{eqapp:defs}
\end{equation}
where $J\bydef\det\bF\,$($=\roO/\ro\,$ due to mass conservation) and
\begin{equation}
\phi(J)\bydef\!-\!J\pE(\roO/J)\,,\quad
\psi(J)\bydef\!-\! J\aE(\roO/J)\,.
\end{equation}
Since the power expended per unit reference volume is equal to $\bS\cdot\dot{\bF}$, prescription \eqref{eqapp:defs} is hyperelastic if and only if the derivative $\DD(\bF)\bydef\partial\bS/\partial\bF$ has the major symmetry for all invertible $\bF$:
\begin{equation}
\big(\DD(\bF)\bL\big)\!\cdot\!\bM = \big(\DD(\bF)\bM\big)\!\cdot\!\bL
\label{eqapp:symcond}
\end{equation}
for all double tensors $\bL,\bM$. To prove that no anisotropic choice -- i.e., $\aE\!\ne\!0\,\Leftrightarrow\,\psi\!\ne\!0$ -- is hyperelastic, it suffices to check condition \eqref{eqapp:symcond} on $\bF=\lambda\,\bI\,$ for all $\lambda>0$, which greatly simplifies calculations. On spherical stretches \eqref{eqapp:symcond} reduces to
\begin{align}
\lambda^3\,\psi'(\lambda^3)\big((\tr\bL)\bM&-(\tr\bM)\bL\big)
\!\cdot\!(\bno\!\tp\bno)
+\,\psi(\lambda^3)\big(\bL\bM - \bM\bL\big)
\!\cdot\!(\bno\!\tp\bno)=0\,.
\label{eqapp:sphcond}
\end{align}
The constitutive map $\phi$ -- and hence $\pE$ -- drops out of \eqref{eqapp:sphcond}, as expected. We now pick $\bL=\bM^{\top}\!\!=\bno\!\tp\bmo\,$, with $\bmo$ a unit vector orthogonal to $\bno$: \mbox{$|\bmo|=1~\&~\bmo\!\cdot\!\bno\!=0\,$}, i.e., two simple shears  in the plane spanned by $\{\bmo,\bno\!\}$, $\bL$ along $\bno$ and $\bM$ along $\bmo$. Since both are traceless and their two products are respectively equal and orthogonal to $\bno\!\tp\bno$, \eqref{eqapp:sphcond} boils down to
\begin{equation}
\psi(\lambda^3)=0\;\;\textrm{for all}\;\;\lambda>0
\quad\Longleftrightarrow\quad\!\aE=0\,.
\label{eqapp:shearcond}
\end{equation}
That $\,\aE\!=\!0\,$ is also sufficient for hyperelasticity is obvious.
\section{\Radc{Nematic hyperelasticity, old and new}}
\label{app:hyperel}
In all part $\Prt$ of a hyperelastic body the power expended equals the time derivative of the stored energy:
\begin{equation}
\label{eqapp:powerid}
\int_{\Prt}\!\!\bT\!\cdot\!(\bna\bv)\,dV\,=
\int_{\Prt}\!\left(\ro\,\sigma\,dV\right)^{\bm{\cdot}}\!,
\end{equation}
%
where $\bT$ is the Cauchy stress, $\bv$ the spatial velocity field, $\ro$ the current mass density, $\sigma$ the stored energy density with respect to mass, and the integration is done with respect to the \emph{current} volume. Since $J\bydef\det\bF\!=\!dV\!/d\Vo$ and the reference volume $\Vo$ does not depend on time, equality \eqref{eqapp:powerid} translates into
\begin{equation}
\label{eqapp:poweridref}
\int_{\Prt}\!\!\left(J\,\bT\,\bF^{-\tsp}\,\right)\!\cdot\dot{\bF}\,d\Vo =
\int_{\Prt}\!(J\ro\,\sigma)^{\bm{\cdot}}\,d\Vo\,,
\end{equation}
%
where use has been made of the differential relation linking the spatial velocity field with the deformation gradient: $\bna\bv=\dot{\bF}\,\bF^{-1}\,$. Mass conservation implies $(J\ro)^{\bm{\cdot}}\!=0\,$. Hence, \eqref{eqapp:poweridref} localizes into
\begin{equation}
\label{eqapp:poweridloc}
\bS\!\cdot\!\dot{\bF} = \roO\dot{\sigma}\,,
\end{equation}
where $\roO\!=\!J\ro$ is the reference mass density and
\begin{equation}
\label{eqapp:Piola}
\bS\bydef J\,\bT\,\bF^{-\tsp}
\end{equation}
is the Piola stress. Since $\dot{\sigma}\!=\!(\partial\sigma/\partial\bF) \!\cdot\!\dot{\bF}$, a necessary and sufficient condition for \eqref{eqapp:powerid} to be satisfied along all motions is that
\begin{equation}
\label{eqapp:hyperPiola}
\bS=\roO\frac{\partial\sigma}{\partial\bF}\,.
\end{equation} \par
On this basis, we now consider the specific constitutive assumption \eqref{eq:We_prerelaxing}--\eqref{eq:trans_strain} and, using \eqref{eq:Ddet}, compute the derivatives
\begin{subequations}
\begin{align}
\frac{\partial\sO}{\partial\bF}&=-\ro\,\sO'\bF^{-\tsp},
\\[1.25ex]
\frac{\partial(\bCtr^{-1}\!\cdot\!\bBo)}{\partial\bF} & =
2\dev(\bCtr^{-1}\,\bBo)\,\bF^{-\tsp}
-\tfrac32(a'/a)(\ro/J)\dev(\bno\!\tp\bno)\!
\cdot\!(\bCtr^{-1}\,\bBo)\,\bF^{-\tsp}.
\end{align}
\end{subequations}
\hfill \linebreak
Summing up all contributions to \eqref{eqapp:hyperPiola} and inverting \eqref{eqapp:Piola} (i.e., calculating $\bT=J^{-1}\bS\,\bF^{\tsp}$) yields \eqref{eq:T_prerelaxing}--\eqref{eq:pressure}. \par
To extend the theory founded on \eqref{eq:We_prerelaxing}--\eqref{eq:trans_strain} to cover the case when the constraint on the nematic texture is lifted and the director is set free to rotate, the stored energy density should be modified at least as follows \cite{XXturzi}:
%
\begin{equation}
\sigmp(\bF,\bn,\!\bna\bn)=
\underbrace{\sigmF(\ro,\bn,\!\bna\bn)}_{\textrm{\normalsize Oseen-Frank}}+\underbrace{\sigmap(\ro,\bn,\bBo)}_{\textrm{\normalsize acoustic}}
\label{eqapp:complete}
\end{equation}
with
\begin{subequations}
\begin{align}
\sigmap(\ro,\bn,\bBo)&=\sO(\ro)
+ \tfrac12\mu(\ro)\tr\!\big(\bCtr^+(\ro,\bn)^{-1}\bBo-\bI\big),
\label{eqapp:acoustic}
\end{align}
\begin{equation}
\bCtr^+(\ro,\bn)\bydef\an(\ro)^2\,\bn\tp\bn+
\an(\ro)^{-1}\big(\,\bI-\bn\tp\bn\big).
\label{eqapp:trans_strain}
\end{equation}
\end{subequations} \par
\Radc{The study of small-amplitude plane waves as done in Sec.\,\ref{sec:elas_sound} and \ref{sec:relax_sound} only depends on the \emph{linearized} features of the theory. Therefore we find it appropriate to provide an explicit expression of the free-energy density function \eqref{eq:We_prerelaxing}--\eqref{eq:trans_strain} -- specialized to a slightly compressible anisotropic fluid as defined in Sec.\,\ref{sec:elas_special} -- when truncated after $O(\eps^2)$ terms, with $\eps$ the smallness parameter reducing the amplitude of the displacement field $\bu$ in \eqref{eq:Ansatz_u}. \par
Let $\bF=\bI+\eps\nabla\bu\,$. Then, the Taylor expansion of the determinant close to the identity yields
\begin{align}
\label{eqapp:J}
J&=\det\bF=\det(\bI+\eps\nabla\bu)
\\
&=1+\eps\tr\!\nabla\bu+
\tfrac12\eps^2\big((\tr\!\nabla\bu)^2-\tr\!\left((\nabla\bu)^2\right)\!\big)
+o(\eps^2)
\notag
\\
&=1+\eps\tr\bE+\tfrac12\eps^2\Big(\!(\tr\bE)^2-\big(\nabla\bu\big)\!\cdot\!\big(\nabla\bu^{\!\top}\big)\!\Big)+o(\eps^2)\,,
\notag
\end{align}
where $\bE\bydef\sym\nabla\bu$ is the infinitesimal deformation. Notably, the differential identity \cite{62Ericksen}
\begin{equation}
\label{eqapp:diffid}
\big(\nabla\bu\big)\!\cdot\!\big(\nabla\bu^{\!\top}\big)=
\divr\!\big((\nabla\bu)\bu-(\divr\bu)\bu\big)+(\tr\bE)^2
\end{equation}
cancels all second-order terms in \eqref{eqapp:J} to within the \emph{null Lagrangian} $-\eps^2n$, with
\begin{equation}
\label{eqapp:nullL}
n\bydef
\tfrac12\divr\!\big((\nabla\bu)\bu-(\divr\bu)\bu\big)\,.
\end{equation}
Therefore, the scaled density variation \eqref{eq:sdvar} expands as
\begin{equation}
\label{eqapp:xi2ord}
\xi=J^{-1}\!-1=
-\eps\tr\bE+\eps^{2}(\tr\bE)^2+\eps^{2}n+o(\eps^2)\,.
\end{equation}
Finally, on account of \eqref{eq:p_iso}, \eqref{eq:p_develop} and \eqref{eq:porder}, the contribution of the isotropic term $\,\sO$ to the elastic energy \emph{per unit reference volume} reads
\begin{equation}
\label{eqapp:ro0s0_develop}
\roO\sO\!=-\eps\,\pO\!\tr\bE+\tfrac12\eps^2\roO\pI\!(\tr\bE)^2\!+o(\eps^2)\,,
\end{equation}
after pruning the ineffective terms, namely, the constant $\roO\sO(\roO)$ and the null Lagrangian $\eps^2\pO n\,$. Also the contribution of the equilibrium pressure $\pO$ to the second-order  term has been dropped, being negligible compared to that of $\roO\pI$\,(cf.\,\eqref{eq:porder}). \par
The isochoric component of the deformation gradient reads
\begin{align}
\bFo&=J^{-1/3}\,\bF=\bI+\eps\bD
-\tfrac13\eps^2\Big((\tr\!\bE)\bD+\tfrac16(\tr\bE)^2\bI
+\tfrac12(\bTh\!\cdot\!\bTh-\bE\!\cdot\!\bE)\bI\Big)+o(\eps^2)\,,
\label{eqapp:Fo}
\end{align}
where
\begin{equation}
\label{eqapp:E&Th&D}
\bTh\bydef\skw\!\nabla\bu\,,\quad
\bD\bydef\dev\!\nabla\bu=\dev\!\bE+\bTh\,,
\end{equation}
with $\skw\!\nabla\bu$ the skew-symmetric part of $\nabla\bu$. The left Cauchy-Green strain tensor associated with $\bFo$ is
\begin{equation}
\begin{split}
\bBo =\bFo\,\bFo^{\top}\! & =\bI+2\eps\dev\!\bE + \eps^2\Big(\!\!-\!\tfrac23(\tr\!\bE)\dev\!\bE-\tfrac19(\tr\bE)^2\bI
+\tfrac13(\bE\!\cdot\!\bE-\bTh\!\cdot\!\bTh)\bI \\
& +(\dev\!\bE+\bTh)(\dev\!\bE-\bTh)\!\Big)+o(\eps^2)\,.
\end{split}
\label{eqapp:Bo_develop}
\end{equation} \par
We now expand the inverse of the right Cauchy-Green transformation strain \eqref{eq:trans_strain} using \eqref{eqapp:xi2ord} and extending \eqref{eq:aaratio} to second order:
\begin{equation}
\label{eqapp:aa2ord}
\an=1+\alO\!+\alI\xi+\tfrac12\alII\xi^2+o(\xi^2)\,.
\end{equation}
As in Sec.~\ref{sec:elas_sound} and \ref{sec:relax_sound}, we will neglect terms of order $O(\alO)$, seen to be insignificant (cf. the last paragraph of Sec.\,\ref{sec:elas_sound}), obtaining
\begin{align}
\bCtr^{-1}&=\bI-3\,\alI\xi\dev(\bno\!\tp\bno)
+3\,\alI^2\xi^2\bno\!\tp\bno - \tfrac32\alII\xi^2\dev(\bno\!\tp\bno)+o(\xi^2) \notag \\[1.ex]
& = \bI+3\,\eps\,\alI\!(\tr\bE)\dev(\bno\!\tp\bno)
+3\,\eps^2\alI^2(\tr\bE)^2\bno\!\tp\bno 
\label{eqapp:Ctrinverse} \\ 
& - 3\,\eps^2\Big(\alI\!n+(\alI\!\!+\tfrac12\alII)(\tr\bE)^2\!\Big)
\!\dev(\bno\!\tp\bno)+o(\eps^2).
\notag
\end{align}
From \eqref{eqapp:Bo_develop} and \eqref{eqapp:Ctrinverse} we finally get the contribution of the anisotropic term in \eqref{eq:We_prerelaxing} to the elastic energy per unit reference volume:
\begin{align}
&\tfrac12\roO\mu\tr\!\big(\bCtr^{-1}\bBo\!-\!\bI\big)=
\roO\muO\eps^2\Big(\!(\dev\bE)\!\cdot\!(\dev\bE)
\label{eqapp:nH_develop}
\\
&~~~~~~+\tfrac92\alI^2(\tr\bE)^2
+3\,\alI(\tr\bE)(\dev\bE)\!\cdot\!(\bno\!\tp\bno)\!\Big)+o(\eps^2).
\notag
\end{align}
Note that all the second-order terms in $\bCtr^{-1}\bBo\!-\!\bI$ affected by either the null Lagrangian $n$ or the infinitesimal rotation $\bTh$ or the second-order coefficient of the anisotropic aspect ratio $\alII$\,are traceless and hence disappear from \eqref{eqapp:nH_develop}. \par
At this point it is appropriate to consider the quadratic free energy surmised by Mullen, L\"{u}thi and Stephen \cite{72mull}, which we alluded to in Sec.\,\ref{sec:intro}: `The experimental anisot\-ropy in the sound velocity [\,\dots] can be explained if at [finite] frequencies a liquid crystal in some respects behaves like a solid and the free energy contains terms like
\begin{equation}
\label{eqapp:72mull}
F=\tfrac12k_{_1}\!(u_{xx}\!+u_{yy})^2
+k_{_2}(u_{xx}\!+u_{yy})u_{zz}+\tfrac12k_{_3}u_{zz}^2,
\end{equation}
where the $k$'s are elastic constants, and the $u_{ij}$ are the elastic strains. We have chosen the $z$ axis to be along the director [\,\dots].' Translated into our component-free notation, \eqref{eqapp:72mull} reads
\begin{equation}
\label{eqapp:72mullR}
F=\tfrac12k_{_1}\!(\trp\bE)^2
+k_{_2}(\trp\bE)\,\epsO
+\tfrac12k_{_3}\epsO^2,
\end{equation}
with
\begin{equation}
\label{eqapp:trp}
\epsO\!\bydef\bE\!\cdot\!(\bno\!\tp\bno)\,,\;\;\;
\trp\bE\bydef\tr\bE-\epsO\,.
\end{equation}
On the other hand, the sum of the quadratic terms in \eqref{eqapp:ro0s0_develop} and \eqref{eqapp:nH_develop} involving the trace of $\bE$ may be reorganized as follows:
\begin{align}
\tfrac12 \roO\pI\!\big(1+\eta\alI\!(9\alI\!\!-2)\big)(\trp\bE)^2\,+
\roO\pI\!\big(1+\eta\alI\!(9\alI\!\!+1)\big)(\trp\bE)\epsO\,+
\tfrac12&\roO\pI\!\big(1+\eta\alI\!(9\alI\!\!+4)\big)\epsO^2
\label{eqapp:SumUp}
\end{align}
where $\eta$ is the small parameter introduced in \eqref{eq:eta}. Therefore, we identify the elastic constants in \eqref{eqapp:72mull} as small perturbations of the bulk modulus $\roO\pI$, parameterized by the product of the shear modulus $\roO\muO$ and the sensitivity coefficient $\alI$. What matters is the slight differences between them, namely,
\begin{equation}
k_{_3}\!-k_{_2}\!=k_{_2}\!-k_{_1}\!=3\roO\muO\alI\!+o(\eta)\,.
\label{eqapp:modiff}
\end{equation}
}
\RADC{\!\!Note that \eqref{eqapp:modiff} implies $k_{_2}^2=k_{_1}k_{_3}\!+o(\eta)$, a condition -- postulated in \cite{72mull} -- that eliminates propagating shear modes at order $O(\eta)$. However, \eqref{eqapp:nH_develop} contains also a term quadratic in $\dev\bE$ of order $O(\eta)$, which sustains such waves.}
\end{document}